\documentstyle[12pt]{article}

\input{epsf.tex}

\begin{document}

\mbox{} \hskip 10cm DF/IST-5.2001

\mbox{} \hskip 10cm May 2001

\vskip 1cm

\begin{center}
{\bf Static and rotating electrically charged black holes in 
three-dimensional Brans-Dicke gravity theories} \\
\vskip 1cm
\'Oscar J. C. Dias\footnote{E-mail: oscar@fisica.ist.utl.pt} \\
\vskip 0.3cm
{\scriptsize  CENTRA, Departamento de F\'{\i}sica,
	      Instituto Superior T\'ecnico,} \\

{\scriptsize  Av. Rovisco Pais 1, 1096 Lisboa, Portugal.} \\
\vskip 0.6cm

Jos\'e P. S. Lemos\footnote{E-mail: lemos@kelvin.ist.utl.pt} \\
\vskip 0.3cm
{\scriptsize  CENTRA, Departamento de F\'{\i}sica,
	      Instituto Superior T\'ecnico,} \\
{\scriptsize  Av. Rovisco Pais 1, 1096 Lisboa, Portugal.} \\

\end{center}

\bigskip

\begin{abstract}
\noindent

We obtain static and rotating electrically charged black holes
of a Einstein-Maxwell-Dilaton theory of the 
Brans-Dicke type in (2+1)-dimensions. The theory is specified 
by three fields, the dilaton $\phi$, the graviton 
$g_{\mu\nu}$ and the electromagnetic field $F^{\mu\nu}$, and 
two parameters, the cosmological constant $\Lambda$ and the 
Brans-Dicke parameter $\omega$. It contains eight different 
cases, of which one distinguishes as special cases, string 
theory, general relativity and a theory equivalent to four 
dimensional general relativity with one Killing vector.
We find the ADM mass, angular momentum, electric charge and
dilaton charge and
compute the Hawking temperature of the solutions. Causal 
structure and geodesic motion of null and timelike particles 
in the black hole geometries are studied in detail.
\newline
\newline
PACS number(s): 04.70.-s, 04.40.-b

\end{abstract}
\newpage
\noindent
{\bf 1. Introduction}
\vskip 3mm
The discovery, by Ba\~nados, Teitelboim and Zanelli 
\cite{btz_PRL,btz_PRD}, of the BTZ black hole solution in 
(2+1)-dimensional Einstein theory with a negative cosmological 
constant has attracted much attention to gravity in (2+1) 
dimensions \cite{Carlip}.
Since then, the inclusion of dilaton or electromagnetic
fields has been made and the corresponding theories have been 
studied.

Einstein-Maxwell theories in (2+1) dimensions have been discussed 
 by several authors \cite{CL1}-\cite{G} who have noticed that if an 
electric charge is included, the BTZ solution presented in the 
original paper \cite{btz_PRL} is valid only in the absence of 
rotation. Rotating electrically charged solutions in the 
Einstein-Maxwell theory have been obtained by Cl\'ement 
\cite{CL1}, and Mart\'{\i}nez, Teitelboim and Zanelli \cite{BTZ_Q} 
using a coordinate transformation. Assuming self dual or anti-self dual
conditions between the electromagnetic fields, Kamata and Koikawa
\cite{KK1} and Cataldo and Salgado \cite{CS} have also obtained
solutions in the Einstein-Maxwell theory. However, as noticed by
Chan \cite{CHAN}, the mass and angular momentum of solutions 
discussed in \cite{KK1} diverge at spatial infinity.  
To overcome this divergence a boundary contribution has been taken
into account \cite{KK2}. The presence of divergences in the 
mass and angular momentum is a usual feature present in charged 
solutions. One way to regularize them is by the introduction of 
a topological Chern-Simons term. This procedure has been done by
Cl\'ement \cite{CL2}, Fernando and  Mansouri \cite{FM} and 
Dereli and Obukhov \cite{DO} who have analyzed the self-dual 
solutions for the Einstein-Maxwell-Chern-Simons theory in (2+1) 
dimensions. 

Einstein-Maxwell-Dilaton theories in (2+1) dimensions have also 
been studied. Chan and Mann \cite{CM} have found static 
electrically charged solutions and Fernando \cite{F} has 
analyzed the Einstein-Maxwell-Dilaton theory when the self dual
 condition is imposed. Chen \cite{CHEN} has applied $T$-duality to 
obtain new rotating
 solutions of Einstein-Maxwell and Einstein-Maxwell-Dilaton 
theories in (2+1) dimensions.

In this paper we find and study in detail the static and rotating
electrically 
charged solutions of a Einstein-Maxwell-Dilaton action of the 
Brans-Dicke type.
The static uncharged solutions were found and analyzed by 
S\'a, Kleber and Lemos \cite{Sa_Lemos_Static} and the angular 
momentum has been added by S\'a and Lemos \cite{Sa_Lemos_Rotat}.
The uncharged theory is specified by two fields, the dilaton 
$\phi$ and the graviton $g_{\mu\nu}$, and two parameters, the 
cosmological constant $\Lambda$ and the Brans-Dicke parameter 
$\omega$. It contains seven different cases and each 
$\omega$ can be viewed as yielding a different dilaton gravity 
theory. For instance, for $\omega=-1$ one gets the simplest 
low-energy string action \cite{CFMP}, for $\omega=0$ one gets a 
theory related (through dimensional reduction) to four 
dimensional General Relativity with one Killing vector 
\cite{Lemos} and for $\omega=\pm \infty$ one obtains three 
dimensional General Relativity analyzed in \cite{btz_PRL,btz_PRD}. 

The electrically charged theory that we are going to study is 
specified by the extra 
electromagnetic field $F^{\mu \nu}$. It contains eight 
different cases. For $\omega=0$ one gets a theory related 
(through dimensional reduction) to electrically charged four 
dimensional General Relativity with one Killing vector 
\cite{Zanchin_Lemos} and for $\omega=\pm \infty$ one obtains 
electrically charged three dimensional General Relativity 
\cite{CL1,BTZ_Q}.

Since magnetically charged solutions in (2+1) dimensions have
totally different properties from the electrically charged ones,
we do not study them here (see \cite{HW}-\cite{YP} for 
magnetically charged solutions of Einstein-Maxwell and 
Einstein-Maxwell-Dilaton theories).

The plan of this article is the following. In Section 2 we set 
up the action and the field equations. The static general 
solution of the field equations are found in section 3 and we 
write the scalar $R_{\mu\nu}R^{\mu\nu} $ which, in (2+1) dimensions,
signals the presence of singularities. The 
angular momentum is added in section 4. In section 5
 we use an extension of the formalism of Regge and Teitelboim 
to derive the mass, angular momentum, electric charge and 
dilaton charge of the 
black holes. In section 6 we study the properties of the  
different cases 
that appear naturally from the solutions. We work out in detail 
the causal structure and the geodesic motion of null and 
timelike particles for typical values of $\omega$ that 
belong to the different ranges. The Hawking temperature is 
computed in section 7. Finally, in section 8 we present the 
concluding remarks.

\vskip 1cm
\noindent
{\bf 2. Field equations}

\vskip 3mm

We are going to work with an action of the 
Maxwell-Brans-Dicke type in three-dimensions written in the 
string frame as
\begin{equation}
S=\frac{1}{2\pi} \int d^3x \sqrt{-g} e^{-2\phi}
  {\bigl [} R - 4 \omega ( \partial \phi)^2
           + \Lambda + F^{\mu \nu}F_{\mu \nu}
  {\bigr ]},                                   \label{ACCAO}
\end{equation}
where
$g$ is the determinant of the 3D metric,
$R$ is the curvature scalar,
$\phi$ is a scalar field called dilaton,
$\Lambda$ is the cosmological constant,
 $\omega$ is the three-dimensional Brans-Dicke parameter and
$F_{\mu\nu} = \partial_\nu A_\mu - \partial_\mu A_\nu$ is 
the Maxwell tensor,
with $A_\mu$ being the vector potential.
Varying this action with respect to $g^{\mu \nu}$, $F^{\mu \nu}$ and
 $\phi$ one gets the Einstein, Maxwell and ditaton equations, 
respectively
\begin{eqnarray}
   & &   \frac{1}{2} G_{\mu \nu}
       -2(\omega+1) \nabla_{\mu}\phi \nabla_{\nu}\phi
       +\nabla_{\mu} \nabla_{\nu}\phi
       -g_{\mu \nu} \nabla_{\gamma} \nabla^{\gamma}\phi   \nonumber \\
   & &  \hskip 4cm
       +(\omega+2) g_{\mu \nu} \nabla_{\gamma}\phi
               \nabla^{\gamma}\phi-\frac{1}{4}g_{\mu \nu}\Lambda=
              \frac{\pi}{2} T_{\mu \nu},
                                        \label{EQUACAO_MET} \\
   & &  \nabla_{\nu}(e^{-2 \phi}F^{\mu \nu})=0 \:,
                                     \label{EQUACAO_MAX} \\
   & &  R -4\omega \nabla_{\gamma} \nabla^{\gamma}\phi
          +4\omega \nabla_{\gamma}\phi \nabla^{\gamma}\phi
          +\Lambda =-F^{\gamma \sigma}F_{\gamma \sigma},
                                         \label{EQUACAO_DIL} 
\end{eqnarray}
where $G_{\mu \nu}=R_{\mu \nu} -\frac{1}{2} g_{\mu \nu} R$ is the
 Einstein tensor, $\nabla$ represents the covariant derivative and 
$T_{\mu \nu}=\frac{2}{\pi}(g^{\gamma \sigma}F_{\mu \gamma}
F_{\nu \sigma}-\frac{1}{4}g_{\mu \nu}F_{\gamma \sigma}
F^{\gamma \sigma})$ is the Maxwell energy-momentum tensor.

We want to consider now a spacetime which is both static and 
rotationally symmetric, implying the existence of a
timelike Killing vector $\partial/\partial t$ and a spacelike 
Killing vector $\partial/\partial\varphi$.
The most general static metric with a Killing vector
$\partial/\partial\varphi$ with closed orbits in three dimensions
can be written as 
$ds^2 = - e^{2\nu(r)} dt^2 + e^{2\mu(r)}dr^2 + r^2 d\varphi^2$, with
$0\leq\varphi\leq 2\pi$. Each different $\omega$ has a very rich and 
non-trivial structure of solutions which could be considered 
on its own. As in \cite{Sa_Lemos_Static,Sa_Lemos_Rotat} we work 
in the Schwarzschild gauge,
$\mu(r)=-\nu(r)$, and compare different black hole solutions in
different theories. For this ansatz the metric is written as
\begin{equation}
  ds^2 = - e^{2\nu(r)} dt^2 + e^{-2\nu(r)}dr^2 + r^2 d\varphi^2.
                               \label{MET_SYM}
\end{equation}
We also assume that the only non-vanishing components of the 
vector potential 
are $A_t(r)$ and $A_{\varphi}(r)$, i.e. , 
\begin{equation}
A=A_tdt+A_{\varphi}d{\varphi}\:.  \label{Potential}
\end{equation}
This implies that the non-vanishing components of the symmetric 
Maxwell tensor
are $F_{tr}$ and $F_{r \varphi}$. 

Inserting the metric (\ref{MET_SYM}) into equation 
(\ref{EQUACAO_MET}) one obtains the following set of 
equations
\begin{eqnarray}
  & &   \phi_{,rr}
      + \phi_{,r} \nu_{,r}
      + \frac{\phi_{,r}}{r}
      - (\omega+2) (\phi_{,r})^2
      - \frac{\nu_{,r}}{2r}
      + \frac{1}{4} \Lambda e^{-2\nu}=
                \frac{\pi}{2}e^{-4\nu} T_{tt} \:,
                                    \label{MET_00}  \\
  & &  - \phi_{,r} \nu_{,r}
       - \frac{\phi_{,r}}{r}
       - \omega (\phi_{,r})^2
       + \frac{\nu_{,r}}{2r}
       - \frac{1}{4} \Lambda e^{-2\nu}=
                  \frac{\pi}{2} T_{rr} \:,
                                    \label{MET_11}  \\
  & &    \phi_{,rr}
       + 2 \phi_{,r} \nu_{,r}
       - (\omega+2) (\phi_{,r})^2
       - \frac{\nu_{,rr}}{2}
       - (\nu_{,r})^2+ \frac{1}{4} \Lambda e^{-2\nu}=
\nonumber \\
  & &  \hskip 8.5cm
              -\frac{\pi}{2}\frac{e^{-2\nu}}{r^2}
                           T_{\varphi \varphi} \:,
                                           \label{MET_22}  \\
  & &   0=\frac{\pi}{2} T_{t \varphi}=
                           e^{-2\nu}F_{tr}F_{\varphi r} \:,
                                           \label{MET_02} 
\end{eqnarray}
where ${}_{,r}$ denotes a derivative with respect to $r$.
In addition, inserting the metric (\ref{MET_SYM}) into equations 
(\ref{EQUACAO_MAX}) and (\ref{EQUACAO_DIL}) yields 
\begin{eqnarray}
 & &  \partial_r {\bigl [} e^{-2 \phi}r(F^{t r}+F^{\varphi r})
{\bigr ]}=0\:,
                                           \label{MAX_0} \\
& &  \omega \phi_{,rr}
      + 2 \omega \phi_{,r} \nu_{,r}
      + \omega \frac{\phi_{,r}}{r}
      - \omega (\phi_{,r})^2
      + \frac{\nu_{,r}}{r}
      + \frac{\nu_{,rr}}{2}
      + (\nu_{,r})^2
      -\frac{1}{4} \Lambda e^{-2\nu}=
\nonumber \\
  & &  \hskip 7cm
      \frac{1}{4}e^{-2\nu}F^{\gamma \sigma}F_{\gamma \sigma} \:.
                                    \label{EQ_DIL}  
\end{eqnarray}

\vskip 1cm

\noindent
{\bf 3. The general static solution}

\vskip 3mm
From the above equations valid for a static and
rotationally symmetric spacetime one sees that equation
(\ref{MET_02}) implies that the electric and
magnetic fields cannot be simultaneously  non-zero, i.e., there 
is no static dyonic solution. In this work we will consider the 
electrically charged case alone ($A_{\varphi}=0,\,A_t\neq 0$).

So, assuming vanishing magnetic field, one has from Maxwell 
equation (\ref{MAX_0}) 
that 
\begin{equation}
F_{tr}=-\frac{\chi}{4r}e^{2 \phi},
\label{MAX_1}
\end{equation}
where $\chi$ is an integration constant which, as we shall see in 
(\ref{CARGA}), is the electric charge. One then has that 
\begin{eqnarray}
 & & F^{\gamma \sigma}F_{\gamma \sigma}=
\frac{\chi^2}{8r^2}e^{4\phi}, \:\:\:\:\:
T_{tt}=\frac{\chi^2}{16 \pi r^2}e^{2\nu}e^{4\phi}, \nonumber \\ 
 & & T_{rr}=-\frac{\chi^2}{16 \pi r^2}e^{-2\nu}e^{4\phi}\:\:,
\:\:\:\:\:T_{\varphi \varphi}=\frac{\chi^2}{16 \pi} \: e^{4\phi}.
\label{MAX_2}
\end{eqnarray}
To proceed we shall first consider the case $\omega\neq -1$.
Adding equations (\ref{MET_00}) and (\ref{MET_11}) one obtains
$\phi_{,rr}=2(\omega+1)(\phi_{,r})^2$,
yielding for the dilaton field the following solution
\begin{equation}
\phi = -\frac{1}{2(\omega+1)}
        \ln [2(\omega+1)r+a_1] + a_2 \:, \quad\quad w \neq -1
                                             \label{DIL}
\end{equation}
where $a_1$ and $a_2$ are constants of integration.
One can, without loss of generality, choose $a_1=0$.
Then, equation (\ref{DIL}) can be written as
\begin{equation}
  e^{-2\phi}= a(\alpha r)^{\frac{1}{\omega+1}}, 
                       \quad\quad w \neq -1 \:,
                                    \label{DILATAO}
\end{equation}
where $\alpha$ is an appropriate constant that is proportional to the 
cosmological constant [see equation (\ref{COSMOL})].
The dimensionless constant $a$ can be viewed
as a normalization to the action (\ref{ACCAO}).
Since it has no influence in our calculations,
apart a possible redefinition of the mass,
we set $a=1$. The vector potential 
$A=A_{\mu}(r)dx^{\mu}=A_t(r)dt$ with 
$A_t(r)=\int F_{tr}dr$ is then
\begin{equation}
A=\frac{1}{4}\chi(\omega+1)
(\alpha r)^{\frac{1}{\omega+1}} dt \:,
                           \quad\quad w \neq -1\:.
                                    \label{VEC_POTENT}
\end{equation}
Inserting the solutions (\ref{MAX_1})-(\ref{DILATAO}) in equations
(\ref{MET_00})-(\ref{EQ_DIL}), we obtain for the metric
\begin{eqnarray}
 ds^2 &=& -{\biggl [}
            (\alpha r)^2 - 
           \frac{b}{(\alpha r)^{\frac{1}{\omega+1}}}
           +\frac{k\chi^2}  
                      {(\alpha r)^{\frac{2}{\omega+1}}} 
            {\biggr ]} dt^2
           + \frac{dr^2}{(\alpha r)^2 - 
           \frac{b}{(\alpha r)^{\frac{1}{\omega+1}}}
           +\frac{k\chi^2}   
                        {(\alpha r)^{\frac{2}{\omega+1}}} }
                                        \nonumber \\
      & &   + r^2 d\varphi^2,
      \hskip 5cm \omega \neq -2,-\frac32,-1,
                                 \label{MET_TODOS}  \\
 ds^2 &=& -{\biggl [} (1+\frac{\chi^2}{4}\ln{r})r^2
                 -br {\biggr ]} dt^2
           +\frac{dr^2}{(1+\frac{\chi^2}{4}
                \ln{r})r^2 -br}
           + r^2 d\varphi^2,                      \nonumber \\
      & &  \hskip 5cm \omega =-2,
                                         \label{MET-2}  \\
 ds^2 &=&  -r^2[-\Lambda \ln (br)+\chi^2r^2] dt^2
           + \frac{dr^2}
            {r^2[-\Lambda \ln (br)+\chi^2r^2]}
           + r^2 d\varphi^2,                      \nonumber \\
      & &  \hskip 5cm \omega=-\frac32,
                                             \label{MET-3/2}
\end{eqnarray}
where $b$ is a constant of integration related with the mass
of the solutions, as will be shown, and 
$k=\frac{(\omega+1)^2}{8(\omega+2)}$.
For $\omega \neq -2,-\frac32,-1 \:$ $\alpha$ is defined as
\begin{equation}
 \alpha =\sqrt{ {\biggl |} \frac{(\omega+1)\Lambda}
     {(\omega+2)(2\omega+3)} {\biggr |}}.
                                 \label{COSMOL}
\end{equation}
For $\omega=-2,-\frac32$ we set $\alpha=1$.
For $\omega=-2$ equations (\ref{MET_00}) and (\ref{MET_11}) imply 
$\Lambda=\chi^2/8$ so, in contrast with the uncharged case 
\cite{Sa_Lemos_Static,Sa_Lemos_Rotat}, the cosmological constant 
is not null.

Now, we consider the case $\omega=-1$. From equations 
(\ref{MET_00})-(\ref{EQ_DIL})
 it follows that $\nu=C_1$,
$\phi=C_2$, where $C_1$ and $C_2$ are constants of integration, 
and that the cosmological constant and electric charge are both 
null, $\Lambda=\chi=0$. So, for $\omega=-1$ the metric gives 
simply the three-dimensional Minkowski spacetime and the dilaton 
is constant,
as occurred in the uncharged case 
\cite{Sa_Lemos_Static,Sa_Lemos_Rotat}.

In (2+1) dimensions, the presence of a curvature 
singularity is revealed by the
scalar $R_{\mu\nu}R^{\mu\nu}$
\begin{eqnarray}
 R_{\mu\nu}R^{\mu\nu}  &=&
       12\alpha^4+ \frac{4 \omega}{(\omega+1)^2}
       \frac{b \alpha^4}
       {(\alpha r)^{\frac{2\omega+3}{\omega+1}}}
     + \frac{(2\omega^2+4\omega+3)}{2(\omega+1)^4}
       \frac{b^2\alpha^4}      
       {(\alpha r)^{\frac{2(2\omega+3)}{\omega+1}}} 
                         \nonumber \\
         & & -\frac{(\omega-1)}{(\omega+1)^2}
       \frac{k \chi^2 \alpha^4}      
       {(\alpha r)^{\frac{2(\omega+2)}{\omega+1}}}
              -\frac{(\omega^2+2\omega+2)}{(\omega+1)^4}
       \frac{k \chi^2 b \alpha^4}      
       {(\alpha r)^{\frac{4\omega+7}{\omega+1}}}
                                           \nonumber \\
         & &    -\frac{(\omega^2+2\omega+3)}{(\omega+1)^4}
       \frac{k^2 \chi^4 \alpha^4}      
       {(\alpha r)^{\frac{4(\omega+2)}{\omega+1}}} \:,
               \hskip 1cm \omega \neq -2,-\frac32,-1 \:,
                                \label{R-2-3/2-1}  \\
  R_{\mu\nu}R^{\mu\nu}  &=&
        8+\frac{32}{r}+\frac{6}{r^2}
        +\chi^2 {\biggl [}6 \ln r+ \frac{4 \ln r}{r} 
        +\frac{3}{r}+5{\biggr ]}+
                                               \nonumber \\
       & &  +\chi^4 {\biggl [}\frac{3}{4}\ln^2 r
         + \frac{5}{4} \ln r +9{\biggr ]} \:,
              \hskip 3cm \omega =-2 \:,
                              \label{R-2}  \\
  R_{\mu\nu}R^{\mu\nu}  &=&
       \Lambda^2 [12\ln^2(br)+20 \ln(br)+9]
       + \Lambda \chi^2 r^2 {\biggr [} 5 \ln(br)+\frac{9}{2}
       {\biggr ]}+\frac{9}{16}\chi^4 r^4  \nonumber \\
       & &                 \hskip 6cm \omega=-\frac32 \:.
                                           \label{R-3/2}
\end{eqnarray}
An inspection of these scalars in (\ref{R-2-3/2-1})-(\ref{R-3/2})
reveals that for $\omega<-2$ and $\omega>-1$ the curvature singularity 
is located at $r=0$ and for $-\frac32<\omega<-1$ the curvature singularity 
is at $r=+\infty$. For $-2 \leq \omega \leq -\frac{3}{2}$ both $r=0$ and 
$r=+\infty$ are singular.
For $\omega=\pm \infty$ spacetime has no curvature singularities. 
Note that in the uncharged case \cite{Sa_Lemos_Static,Sa_Lemos_Rotat},
for $-2 \leq \omega < -\frac{3}{2}$ the curvature singularity is located 
only at $r=0$.

%

\vskip 2.5cm

\noindent
{\bf 4. The general rotating solution}

\vskip 3mm

In order to add angular momentum to the spacetime we perform
the following coordinate transformations (see e.g. 
\cite{Sa_Lemos_Rotat}-\cite{HorWel})
\begin{eqnarray}
 t &\mapsto& \gamma t-\frac{\theta}{\alpha^2} \varphi \:,
                                       \nonumber  \\
 \varphi &\mapsto& \gamma \varphi-\theta t \:,
                                       \label{TRANSF_J}
\end{eqnarray}
where $\gamma$ and $\theta$ are constant parameters.
Substituting (\ref{TRANSF_J}) into
(\ref{MET_TODOS})-(\ref{MET-3/2}) we obtain
\begin{eqnarray}
 ds^2 &=& -{\biggl [} {\biggl (}\gamma^2-   
           \frac{\theta^2}{\alpha^2} {\biggr )}
            (\alpha r)^2 - 
     \frac{\gamma^2 b}{(\alpha r)^{\frac{1}{\omega+1}}}
           + \frac{\gamma^2 k\chi^2}     
                      {(\alpha r)^{\frac{2}{\omega+1}}} 
            {\biggr ]} dt^2       \nonumber \\
      & &
           -\frac{\gamma \theta}{\alpha^2}{\biggl [}
            \frac{b}
            {(\alpha r)^{\frac{1}{\omega+1}}}
           -\frac{k\chi^2}{(\alpha r)^    
         {\frac{2}{\omega+1}}}  {\biggr ]} 2dt d\varphi    
            + \frac{dr^2}{(\alpha r)^2 - 
           \frac{b}{(\alpha r)^{\frac{1}{\omega+1}}}
           +\frac{k\chi^2}        
                           {(\alpha r)^{\frac{2}{\omega+1}}} }
                                        \nonumber \\
      & &   + {\biggl [} {\biggl (}\gamma^2-   
           \frac{\theta^2}{\alpha^2} {\biggr )}r^2 + 
           \frac{\theta^2}{\alpha^4}\frac{b}
           {(\alpha r)^{\frac{1}{\omega+1}}}
           - \frac{\theta^2}{\alpha^4}\frac{k\chi^2}          
                         {(\alpha r)^{\frac{2}{\omega+1}}} 
            {\biggr ]} d\varphi^2,           \nonumber \\
      & &
      \hskip 7cm \omega \neq -2,-\frac32,-1,
                                 \label{MET_TODOS_J}  \\
 ds^2 &=&\!\! -{\biggl [} {\biggl (}(\gamma^2- \theta^2)+   
                  \frac{\gamma^2 \chi^2}{4}\ln{r} {\biggr )} 
            r^2    -\gamma^2 b r {\biggr ]} dt^2
           + \gamma \theta {\biggl [}
           \frac{\chi^2}{4}r^2\ln{r}
           - b r {\biggr ]} 2 dt d\varphi        \nonumber \\
      & &
           +\frac{dr^2}{(1+\frac{\chi^2}{4}
                \ln{r})r^2 -br}
           + {\biggl [} {\biggl (}(\gamma^2- \theta^2)-        
              \frac{\theta^2 \chi^2}{4}\ln{r} 
           {\biggr )}r^2 +\theta^2 b r {\biggr ]} d\varphi^2,         
                                                 \nonumber \\
      & &  \hskip 5cm \omega =-2,
                                         \label{MET-2_J}  \\
 ds^2 &=&\!\!  -r^2[-\gamma^2 \Lambda \ln(br)-\theta^2+ 
                        \gamma^2 \chi^2r^2] dt^2
           -\gamma \theta r^2[\Lambda \ln(br)+1-
            \chi^2r^2] 2dt d\varphi      \nonumber \\
      & &
        +\frac{dr^2}{r^2[-\Lambda \ln (br)+ \chi^2r^2]}
           + r^2[\theta^2 \Lambda \ln(br)+\gamma^2            
           -\theta^2 \chi^2r^2] d\varphi^2,                           
                                             \nonumber \\
      & &  \hskip 5cm \omega=-\frac32 \:.
                                             \label{MET-3/2_J}
\end{eqnarray}
Introducing transformations (\ref{TRANSF_J}) into
(\ref{VEC_POTENT}) we obtain that the vector potential 
$A=A_{\mu}(r)dx^{\mu}$ is now given by
\begin{equation}
A=\gamma A(r)dt-
\frac{\theta}{\alpha^2}A(r) d\varphi\:,
                           \quad\quad w \neq -1\:,
                                    \label{VEC_POTENT_J}
\end{equation}
where $A(r)=\frac{1}{4}\chi(\omega+1)(\alpha r)^{\frac{1}{\omega+1}}$.
Solutions (\ref{MET_TODOS_J})-(\ref{VEC_POTENT_J}) represent
electrically charged stationary spacetimes and also solve 
(\ref{ACCAO}). Analyzing the Einstein-Rosen bridge of the 
static solution one concludes that spacetime is not simply 
connected which implies that the first Betti number of the manifold is 
one, i.e., closed curves encircling the horizon cannot be shrunk 
to a point. So, transformations (\ref{TRANSF_J}) generate a new metric
because they are not permitted global coordinate transformations
 \cite{Stachel}. Transformations 
(\ref{TRANSF_J}) can be done locally, but not globally. 
Therefore metrics (\ref{MET_TODOS})-(\ref{MET-3/2}) and 
(\ref{MET_TODOS_J})-(\ref{MET-3/2_J}) can be locally mapped into 
each other but not globally, and such they are distinct.

\vskip 1cm

\noindent
{\bf 5. Mass, angular momentum and electric charge of the solutions}

\vskip 3mm

In this section we will calculate the mass, angular momentum, 
electric charge and dilaton charge of the static and rotating 
electrically charged 
black hole solutions. To obtain these quantities  
we apply the formalism  of Regge and Teitelboim \cite{Regge}
(see also \cite{BTZ_Q,Sa_Lemos_Static,Sa_Lemos_Rotat,Lemos}).

We first write the metrics (\ref{MET_TODOS_J})-(\ref{MET-3/2_J}) 
in the canonical form involving the lapse function $N^0(r)$ and 
the shift function $N^{\varphi}(r)$
\begin{equation}
     ds^2 = - (N^0)^2 dt^2
            + \frac{dr^2}{f^2}
            + H^2(d\varphi+N^{\varphi}dt)^2 \:,
                               \label{MET_CANON}
\end{equation}
where $f^{-2}=g_{rr}$, $H^2=g_{\varphi \varphi}$,
 $H^2 N^{\varphi}=g_{t \varphi}$ and $(N^0)^2-H^2(N^{\varphi})^2=g_{tt}$.
Then, the action can be written in the hamiltonian form as a 
function of the energy constraint ${\cal{H}}$, momentum constraint 
${\cal{H}}_{\varphi}$ and Gauss constraint $G$
\begin{eqnarray}
S &=& -\int dt d^2x[N^0 {\cal{H}}+N^{\varphi} {\cal{H}_{\varphi}}+ A_{t} G]+ 
                     {\cal{B}}          \nonumber \\
 &=&  -\Delta t \int dr N
        {\biggl [} \frac{2 \pi^2}{H^3}e^{-2 \phi}-
        4f^2(H \phi_{,r}e^{-2 \phi})_{,r}-2H \phi_{,r}(f^2)_{,r}
        e^{-2 \phi}                                        \nonumber \\
 & &    +2f(fH_{,r})_{,r}e^{-2 \phi}+4 \omega H f^2
        (\phi_{,r})^2e^{-2 \phi}-\Lambda H e^{-2 \phi}+
\frac{2H}{f}e^{-2 \phi}(E^2+B^2){\biggr ]}
\nonumber \\
 & & 
      + \Delta t \int dr N^{\varphi}{\biggl [}{\bigl (}2 \pi e^{-2 \phi}
       {\bigr )}_{,r}+\frac{4H}{f}e^{-2 \phi}E^rB{\biggr ]} \nonumber \\
 & &   
       + \Delta t \int dr A_t {\biggl [}-\frac{4H}{f}
       e^{-2 \phi} \partial_r E^r{\biggr ]} +{\cal{B}} \:,
                               \label{ACCAO_CANON}
\end{eqnarray}
where $N=\frac{N^0}{f}$, 
$\pi \equiv {\pi_{\varphi}}^r=-\frac{fH^3 (N^{\varphi})_{,r}}{2N^0}$ 
(with $\pi^{r \varphi}$ being the momentum conjugate to 
$g_{r \varphi}$),  $E^r$ and $B$ are the electric and magnetic fields 
and ${\cal{B}}$ is a boundary term.
Upon varying the action with respect to $f(r)$, $H(r)$, $\pi(r)$, 
$\phi(r)$ and $E^r(r)$ one picks up additional surface terms.
Indeed,
\begin{eqnarray}
\delta S &=& - \Delta t N {\biggl [}(H_{,r}-2H\phi_{,r})e^{-2\phi}
         \delta f^2 -(f^2)_{,r}e^{-2\phi}\delta H -4f^2 H 
             e^{-2\phi} \delta(\phi_{,r})
                                               \nonumber \\
         & &              
         +2H{\bigl [}(f^2)_{,r}+4(\omega+1)f^2 \phi_{,r}{\bigr ]}
         e^{-2\phi}\delta \phi
                +2f^2 e^{-2\phi}\delta (H_{,r}) {\biggr ]}
                                        \nonumber \\
         & & +\Delta t N^{\varphi}{\biggl [}2e^{-2\phi}\delta \pi
         -4 \pi e^{-2\phi}\delta \phi {\biggr ]}+ \Delta t A_t
{\biggl [}- \frac{4H}{f}e^{-2 \phi} \delta E^r{\biggr ]}         
+ \delta {\cal{B}}
          \nonumber \\
         & & +(\mbox{terms vanishing when the
                    equations of motion hold}).
                               \label{DELTA_ACCAO}
\end{eqnarray}
In order that the Hamilton's equations are satisfied,
the boundary term ${\cal{B}}$ has to be adjusted so that it cancels 
the above additional surface terms. More specifically one has
\begin{equation}
  \delta {\cal{B}} = -\Delta t N \delta M +\Delta t N^{\varphi}\delta J+
             \Delta t A_t \delta Q \:,
                              \label{DELTA_B}
\end{equation}
where one identifies $M$ as the mass, $J$ as the angular momentum and
$Q$ as the electric charge since they are the terms conjugate to the 
asymptotic values of $N$, $N^{\varphi}$ and $A_t$, respectively.

To determine the $M$, $J$ and $Q$ of the black hole one must take the
black hole spacetime and subtract the background
reference spacetime contribution, i.e., we choose the energy zero point 
in such a way that the mass, angular momentum and charge vanish
when the black hole is not present.

Now, note that for 
$\omega <-2$, $\omega>-3/2$ and $\omega \neq -1$, spacetime
(\ref{MET_TODOS_J}) has an asymptotic metric given by
\begin{equation}
-{\biggl (}\gamma^2-\frac{\theta^2}{\alpha^2} {\biggr )}
 \alpha^2 r^2 dt^2+ \frac{d r^2}{ \alpha^2 r^2}+
{\biggl (}\gamma^2-\frac{\theta^2}{\alpha^2} {\biggr )}
 r^2 d \varphi^2 \:,
                                          \label{ANTI_SITTER}
\end{equation}
i.e., it is asymptotically an anti-de Sitter spacetime.
In order to have the usual form of the anti-de Sitter metric we 
choose $\gamma^2-\theta^2 / \alpha^2=1$. For the cases 
$-2\leq \omega \leq -3/2$ we shall also choose 
$\gamma^2-\theta^2 / \alpha^2=1$, as has been done
for the uncharged case \cite{Sa_Lemos_Static,Sa_Lemos_Rotat}. 
For $\omega \neq -3/2,-1$ the anti-de Sitter spacetime is 
also the background reference spacetime, since the metrics  
(\ref{MET_TODOS_J}) and (\ref{MET-2_J})
 reduce to (\ref{ANTI_SITTER}) if the
black hole is not present ($b=0$ and $\varepsilon=0$). 
For $\omega=-3/2$ the above described procedure of choosing 
the energy zero point does not apply since for any value 
of $b$ and $\varepsilon$ one still has a black hole solution.
Thus, for $\omega=-3/2$ the energy zero point is chosen
arbitrarily to correspond to the black hole solution with 
$b=1$ and $\varepsilon=0$.

Taking the subtraction of the background reference spacetime 
into account and noting that $\phi-\phi_{\rm ref}=0$ and that
$\phi_{,r}-\phi_{,r}^{\rm ref}=0$ we have that the mass, 
angular momentum and electric charge are given by
\begin{eqnarray}
M &=& (2H\phi_{,r}-H_{,r})e^{-2\phi}(f^2-f^2_{\rm ref})
      +(f^2)_{,r}e^{-2\phi}(H-H_{\rm ref}) \nonumber \\
 & &
      -2f^2 e^{-2\phi}(H_{,r}-H_{,r}^{\rm ref})\:,  
                                            \nonumber \\
J &=&  -2e^{-2\phi} (\pi-\pi_{\rm ref}) \:,
                                            \nonumber \\
Q &=&  \frac{4H}{f}e^{-2 \phi} (E^r-E^r_{\rm ref}) \:.         
                                    \label{MJQ_GERAL}
\end{eqnarray}  
Then, for $\omega>-3/2$  and $\omega \neq -1$, we finally have
that the mass and angular momentum are (after taking the 
appropriate asymptotic limit: $r \rightarrow +\infty$ for 
$\omega>-1$ and $r \rightarrow 0$ for $-3/2<\omega<-1$, see 
the Penrose diagrams on section 6.3 to understand the reason 
for these limits)
\begin{eqnarray}
M &=& b {\biggl [}\frac{\omega +2}{\omega +1}\gamma^2
  + \frac{\theta^2}{\alpha^2}{\biggl ]}=M_{{\rm Q}=0} \:,  
    \nonumber \\
J &=&  \frac{\gamma \theta}{\alpha^2}b\frac{2\omega+3} 
{\omega +1}=J_{{\rm Q}=0}\:,   \hskip 2cm \omega> -3/2, 
\: \omega \neq -1  \:,    
                                    \label{MJ>-1}
\end{eqnarray}  
where $M_{{\rm Q}=0}$ and $J_{{\rm Q}=0}$ are the mass and 
angular momentum of the uncharged black hole.
For $\omega<-3/2$, the mass and angular momentum are 
(after taking the appropriate asymptotic limit, $r \rightarrow +\infty$)
\begin{eqnarray}
M &=& b {\biggl [}\frac{\omega +2}{\omega +1}\gamma^2
  + \frac{\theta^2}{\alpha^2}{\biggl ]} 
  + {\rm Div_M}(\chi,r)= M_{{\rm Q}=0}
  + {\rm Div_M}(\chi,r) \:,
            \label{M<-1}   \\
J &=&  \frac{\gamma \theta}{\alpha^2}b\frac{2\omega+3} {\omega +1} 
+ {\rm Div_J}(\chi,r)=
J_{{\rm Q}=0}+ {\rm Div_J}(\chi,r)
\:,   \nonumber \\
& & \hskip 6cm \omega<-3/2 \:,    
                                    \label{J<-1}
\end{eqnarray}  
where ${\rm Div_M}(\chi,r)$ and ${\rm Div_J}(\chi,r)$ 
are terms proportional to the charge $\chi$ that diverge at the asymptotic 
limit. These kind of  divergent terms are present even in the case 
$\omega=\pm \infty$ which gives the electrically charged BTZ black 
hole \cite{BTZ_Q}. Following \cite{BTZ_Q} these divergences can be 
treated as follows. One considers a boundary of large radius $r_0$
involving the black hole. Then, one sums and subtracts 
${\rm Div_M}(\chi,r_0)$ to (\ref{M<-1}) so that the mass 
(\ref{M<-1}) is now written as 
\begin{equation}
M = M(r_0)+ [{\rm Div_M}(\chi,r)-
     {\rm Div_M}(\chi,r_0)] \:,
           \label{M0<-1}
\end{equation}  
where $M(r_0)=M_{{\rm Q}=0}+{\rm Div_M}(\chi,r_0)$, i.e.,
\begin{equation}
M_{{\rm Q}=0}=M(r_0)-
{\rm Div_M}(\chi,r_0)\:.
                       \label{M0_v2<-1}
\end{equation}  
The term between brackets in (\ref{M0<-1}) vanishes when 
$r\rightarrow r_0$. Then $M(r_0)$ is the energy within the 
radius $r_0$. The difference between $M(r_0)$ and 
$M_{{\rm Q}=0}$ is $-{\rm Div_M}(\chi,r_0)$ which is 
interpreted as the electromagnetic energy outside $r_0$ 
apart from an infinite constant which is absorbed in 
$M(r_0)$. The sum (\ref{M0_v2<-1}) is then independent of 
$r_0$, finite and equal to the total mass.

To handle the angular momentum divergence, one first notice 
that the asymptotic limit of the angular momentum per unit  mass 
$(J/M)$ is either zero or one, so the angular momentum diverges at 
a rate slower or equal to the rate of the mass divergence. 
The divergence on the angular momentum can then be treated 
in a similar way as the mass divergence. So, the divergent 
term $-{\rm Div_J}(\chi,r_0)$ can be interpreted as 
the electromagnetic angular momentum outside $r_0$ up to an 
infinite constant that is absorbed in $J(r_0)$.

In practice the treatment of the mass and angular divergences 
amounts to forgetting about $r_0$ and take as zero the asymptotic 
limits: $\lim {\rm Div_M}(\chi,r)=0$ and 
$\lim {\rm Div_J}(\chi,r)=0$. So, for $\omega<-3/2$ 
the mass and angular momentum are also 
given by (\ref{MJ>-1}).

Interesting enough, as has been noticed in \cite{BTZ_Q}, is 
the fact that in four spacetime dimensions there occurs a 
similar situation. For example, the $g_{tt}$ component of 
Reissner-Nordstr\"{o}m solution can be written as
$1-\frac{2M(r_0)}{r}+Q^2(\frac{1}{r}-\frac{1}{r_0})$. The 
total mass $M=M(r_0)+\frac{Q^2}{2r_0}$ 
is independent of $r_0$ and 
$\frac{Q^2}{2r_0}$ is the electrostatic
energy outside a sphere of radius $r_0$. 
In this case, since $\frac{Q^2}{2r_0}$ vanishes when 
$r_0 \rightarrow \infty$, one does not need to include an 
infinite constant in $M(r_0)$. 
Thus, in this general Brans-Dicke theory in 3D we conclude 
that both situations can occur depending on the value of 
$\omega$. The $\omega>-3/2, \:\omega \neq -1$ case is 
analogous to the the Reissner-Nordstr\"{o}m black hole in 
the sense that it is not necessary to include an infinite 
constant in $M(r_0)$, while the case 
$\omega<-3/2$ is similar to the 
electrically charged BTZ black hole \cite{BTZ_Q} since  an 
infinite constant must be included in  $M(r_0)$.

For $\omega=-3/2$ the mass and angular momentum are 
ill defined since the boundaries $r \rightarrow \infty$
and $r \rightarrow 0$ have logarithmic singularities 
in the mass term even in the absence of the electric charge.

Now, we calculate the electric charge of the black holes. 
To determine the electric field we must consider the 
projections of the Maxwell field on spatial hypersurfaces. 
The normal to such hypersurfaces is 
$n^{\nu}=(1/N^0,0,-N^{\varphi}/N^0)$ so the electric field is
$E^{\mu}=g^{\mu \sigma}F_{\sigma \nu}n^{\nu}$. Then, from 
(\ref{MJQ_GERAL}), the electric charge is
\begin{equation}
 Q=-\frac{4Hf}{N^0}e^{-2 \phi}(\partial_rA_t-N^{\varphi}
    \partial_r A_{\varphi})=\gamma \chi\:,     
\hskip 2cm  \omega \neq -1 \:.
\label{CARGA}
\end{equation}

One can also define a dilaton charge as the flux
of the  dilaton field over an asymptotic one-sphere
\begin{equation}
Q_{\rm Dil}\equiv \frac{2}{\pi}\int d \varphi \sqrt{\sigma} u^{\mu} 
\partial_{\mu}(\phi-\phi_{\rm {ref}}) \:,
\label{CARGA_DIL}
\end{equation}
where $\sqrt{\sigma}=\sqrt{g_{\varphi \varphi}}$ and 
$u^{\mu}=(0,\sqrt{g^{rr}},0)$ is the normal vector to
the one-sphere boundary. As the background reference 
spacetime has the same dilaton solution as the black hole spacetime 
one conclude that the dilaton charge is zero. 

The mass, angular momentum and electric charge of the static 
black holes can be obtained by putting $\gamma=1$ and 
$\theta=0$ on the above expressions [see (\ref{TRANSF_J})].

Now, we want to cast the metric in terms of $M$, $J$ and $Q$.
For $\omega \neq -2,-3/2,-1$, we can use (\ref{MJ>-1}) to 
solve a quadratic equation for 
$\gamma^2$ and $\theta^2 / \alpha^2$. It gives two distinct 
sets of solutions
\begin{equation}
\gamma^2=\frac{\omega+1}{2(\omega+2)}\frac{M(2- \Omega)}{b} \:,\:\:\:
\:\:\:\: \frac{\theta^2}{\alpha^2}=\frac{M \Omega}{2b}\:, 
\label{DUAS}
\end{equation}

\begin{equation}
\gamma^2=\frac{\omega+1}{2(\omega+2)}\frac{M \Omega}{b} \:,\:\:\:
\:\:\:\: \frac{\theta^2}{\alpha^2}=\frac{M(2- \Omega)}{2b}\:, 
\label{DUAS_ERR}
\end{equation}
where we have defined a rotating parameter $\Omega$ as
\begin{equation}
\Omega \equiv 1- \sqrt{1-\frac{4(\omega+1)(\omega+2)}
{(2\omega+3)^2}\frac{J^2 \alpha^2}{M^2}} \:,      
\hskip 1cm  \omega \neq -2,-3/2,-1 \:.
                    \label{OMEGA}
\end{equation}
When we take $J=0$ (which implies $\Omega=0$), 
(\ref{DUAS}) gives $\gamma \neq 0$ and $\theta= 0$ while
(\ref{DUAS_ERR}) gives the nonphysical solution $\gamma=0$ and 
$\theta \neq 0$ which does not reduce to the static original metric. 
Therefore
we will study the solutions found from (\ref{DUAS}).
For $\omega=-2$ we have $\gamma^2=J^2/Mb$ and $\theta^2=M/b$.

The condition that $\Omega$ remains real imposes for 
$-2>\omega>-1$ a restriction on the allowed values of the 
angular momentum:
$|\alpha J|\leq \frac{|2\omega+3|M}{2\sqrt{(\omega+1)
(\omega+2)}}$. For $-2>\omega>-1$ we have 
$0 \leq \Omega \leq 1$. In the range $-2<\omega<-3/2$ 
and $-3/2<\omega<-1$ we have $\Omega <0$. The condition 
$\gamma^2-\theta^2/\alpha^2=1$ fixes the value of $b$ 
and from (\ref{CARGA}) we can write $k \chi^2$ 
as a function of $b,M,\Omega,Q$. Thus,
\begin{eqnarray}
b &=&  \frac{M}{2(\omega +2)}{\biggl [}2(\omega+1)-
       (2\omega +3)\Omega  {\biggr ]} \:,
                                \label{b}  \\ 
k\chi^2&=&  \frac{b}{4}(\omega +1) \frac{Q^2}
                    {M(2-\Omega)}\:,
\hskip 2cm  \omega \neq -2,-3/2,-1 \:,    
                                    \label{c^2}
\end{eqnarray}
and
\begin{eqnarray}
b = \frac{J^2-M^2}{M}\:,\:\: \:\:\:\:\:\: 
\chi^2 =  \frac{Q^2 M b}{J^2}\:,                               
\hskip 2cm  \omega=-2 \:.
\label{bc_2}
\end{eqnarray}
The metrics (\ref{MET_TODOS_J}) and (\ref{MET-2_J}) may 
now be cast in the form
\begin{eqnarray}
 ds^2 &=& -{\biggl [} (\alpha r)^2 - \frac{(\omega+1)}
          {2(\omega+2)} \frac{M(2-\Omega)}
          {(\alpha r)^{\frac{1}{\omega+1}}}
           + \frac{(\omega+1)^2}{8(\omega+2)} 
          \frac{Q^2} {(\alpha r)^{\frac{2}{\omega+1}}} 
            {\biggr ]} dt^2       \nonumber \\
      & &
          -\frac{\omega+1}{2\omega+3}J{\biggl [}
            (\alpha r)^{-\frac{1}{\omega+1}}
           -\frac{(\omega+1)Q^2}{4M(2-\Omega)}
           (\alpha r)^{-\frac{2}{\omega+1}}  {\biggr ]} 
               2dt d\varphi 
                                      \nonumber \\
      & &       
           + {\biggl [}(\alpha r)^2 - 
           \frac{M[2(\omega+1)-
       (2\omega +3)\Omega]}{2(\omega +2)
            (\alpha r)^{\frac{1}{\omega+1}}}
                                       \nonumber \\
      & &  \:\:\:\:\:\:\:\:\:\:  
       +\frac{(\omega +1)Q^2[2(\omega+1)-
       (2\omega +3)\Omega]}
           {8(\omega +2)(2-\Omega)(\alpha r)^{\frac{2}
           {\omega+1}}} {\biggr ]}^{-1} dr^2 
                                        \nonumber \\
      & &   + \frac{1}{\alpha^2}{\biggl [} (\alpha r)^2 + 
           \frac{M \Omega}
           {2(\alpha r)^{\frac{1}{\omega+1}}}
           - \frac{(\omega +1) \: \Omega \: Q^2} {8(2-\Omega)
            (\alpha r)^{\frac{2}{\omega+1}}} 
            {\biggr ]} d\varphi^2,           \nonumber \\
      & &
      \hskip 7cm \omega \neq -2,-\frac32,-1\:,
                                        \label{MET_MJQ} \\
 ds^2 &=& -{\biggl [}{\biggl (}1+\frac{Q^2}{4}\ln r{\biggl )}r^2 
          -\frac{J^2}{M}\: r{\biggr ]} dt^2      
          +J{\biggl [} \frac{Q^2 M}{4 J^2}r^2 \ln r-r
                {\biggr ]}  2dt d\varphi 
                                       \nonumber \\
      & &       
           + {\biggl [} {\biggl (}1+\frac{Q^2}{4}
          {\bigl (}1-\frac{M^2}{J^2}{\bigl )}\ln r{\biggl )}r^2 
          -M {\biggl (}\frac{J^2}{M^2}-1 {\biggl )}r 
          {\biggr ]}^{-1} dr^2 
                                        \nonumber \\
      & & + {\biggl [}{\biggl (}1-\frac{Q^2 M^2}{4 J^2}\ln r
          {\biggl )}r^2+Mr
            {\biggr ]} d\varphi^2,   
      \hskip 2cm \omega = -2.
                                        \label{MET_MJQ_-2} 
\end{eqnarray}
Analyzing the function $\Delta=g_{rr}^{-1}$ in 
(\ref{MET_TODOS_J}), (\ref{MET-2_J}), and (\ref{b})-(\ref{bc_2}) we can set 
the conditions imposed on the mass and angular momentum of 
the solutions obtained for the different values of 
$\omega \neq -3/2,-1$, in order that the black holes might
 exist. These conditions are summarized on Table 1.

\vfill
\vskip 5mm
\noindent
\begin{tabular}{|l|l|l|}    \hline 
     Range of $\omega$  
        & Black holes with $M>0$ 
        & Black holes with $M<0$      \\ \hline\hline
     $-\infty<\omega<-2$   
        &  $|\alpha J|\leq\frac{|2\omega+3|M}{2\sqrt{(\omega+1)
                                              (\omega+2)}}$  
        &  $|\alpha J|\leq\frac{|(2\omega+3)M|}{2\sqrt{(\omega+1)
                                              (\omega+2)}}$  
                                       \\ \hline
     $\omega=-2$     
        &  might exist for any $J$ 
        &  $|J|<|M|$                    \\ \hline
     $-2<\omega<-\frac32$     
        &  do not exist for any $J$
        &  might exist for any  $J$               \\ \hline
     $-\frac32<\omega<-1$       
        &  $|\alpha J|>M$
        &  might exist for any $J$   \\ \hline
     $-1<\omega<+\infty$         
        & $|\alpha J| \leq
                   \frac{(2\omega+3)M}{2\sqrt{(\omega+1)
                                         (\omega+2)}}$               
        & $|\alpha J|  \leq \frac{(2\omega+3)|M|}                         
                        {2\sqrt{(\omega+1)(\omega+2)}}$ 	
		                              \\ \hline
\end{tabular}
\vskip 1mm
{\small
\noindent
{\bf Table 1.}
\noindent
Values of the angular momentum for which  black holes with 
positive and negative masses might exist.
}
\vskip 3mm

We can mention the principal differences between the 
charged and uncharged theory. The charged theory has 
black holes which are not present in the uncharged theory 
for the following range of parameters: (i)  
$-\infty<\omega<-2$, $M<0$; (ii) $\omega=-2$, $M>0$, $|J|<M$; 
(iii) $-3/2<\omega<-1$, $M<0$, $|\alpha J|>M$; (iv)
 $-1<\omega<+\infty$, $M>0$, $M<|\alpha J|<
(2\omega+3)M/2\sqrt{(\omega+1)(\omega+2)}$ and (v)  
$-1<\omega<+\infty$, $M<0$, $|\alpha J|<|M|$.
\vskip 1cm

\noindent
{\bf 6. Causal and geodesic structure of the charged black holes}

\vskip 3mm
\noindent
{\bf 6.1. Analysis of the causal structure}

\vskip 3mm
In order to study the causal structure we follow the procedure 
of Boyer and Lindquist \cite{Boyer} and Carter \cite{Carter} 
and write the metrics (\ref{MET_TODOS_J})-(\ref{MET-3/2_J}) 
in the form [see (\ref{TRANSF_J})]
\begin{equation}
ds^2=-\Delta {\biggl (}\gamma dt-\frac{\theta}{\alpha^2} 
      d\varphi {\biggr )}^2+\frac{dr^2}{\Delta} 
      +r^2{\biggr (}\gamma d\varphi-\theta dt{\biggr )}^2 \:,
                                    \label{MET_PENR}
\end{equation}
where
\begin{eqnarray}
 \Delta &=& 
            (\alpha r)^2 - 
           b(\alpha r)^{-\frac{1}{\omega+1}}
           +k\chi^2(\alpha r)^{-\frac{2}{\omega+1}} \:,
          \hskip 1cm \omega \neq -2,-3/2,-1, 
                                 \label{DELTA_TODOS}  \\
 \Delta &=&  (1+\frac{\chi^2}{4}\ln{r})r^2  -br \:,    
                     \hskip 3cm \omega=-2,
                                 \label{DELTA-2}  \\
\Delta &=&  r^2[-\Lambda \ln (br)+\chi^2r^2] \:,
               \hskip 3cm \omega=-\frac32\:,
                                             \label{DELTA-3/2}
\end{eqnarray}
and in (\ref{DELTA_TODOS}) $b$ and $k\chi^2$ are 
given by (\ref{b}) and (\ref{c^2}). 

Bellow we describe the general procedure to draw the 
Penrose diagrams. Following Boyer and Lindquist \cite{Boyer}, 
we choose a new angular coordinate which straightens out 
the spiraling null geodesics that pile up around the event 
horizon. A good choice is 
\begin{equation}
\bar{\varphi}=\gamma \varphi-\theta t\:.
                                    \label{TRANSF_RECTA}
\end{equation}
Then (\ref{MET_PENR}) can be written as
\begin{equation}
ds^2=-\Delta {\biggl (} \frac{1}{\gamma} dt-\frac{\theta}
   {\alpha^2 \gamma} d \bar{\varphi} {\biggr )}^2
   +\frac{dr^2}{\Delta} +r^2 d \bar{\varphi}^2 \:.
                                    \label{MET_PENR_RECTA}
\end{equation}
Now the null radial geodesics
are straight lines at $45^{\rm o}$.
The advanced and retarded null coordinates are defined by
\begin{equation}
u=\gamma t-r_{\ast}\:, \hskip 1cm
v=\gamma t+r_{\ast}\:,
                                    \label{NULL_COORD}
\end{equation}
where $r_{\ast}=\int\Delta^{-1}dr$ is the tortoise coordinate. 
In general, the integral defining the tortoise coordinate 
cannot be solved explicitly for the solutions 
(\ref{DELTA_TODOS})-(\ref{DELTA-3/2}). Moreover, the maximal 
analytical extension depends critically on the values of 
$\omega$. There are seven cases which have to be treated 
separately: $\omega<-2$, $\omega=-2$, $-2<\omega<-3/2$, 
$\omega=-3/2$, $-3/2<\omega<-1$, $\omega>-1$ and 
$\omega=\pm \infty$. As we shall see, on some of the cases 
the $\Delta$ function has only one zero and so the black 
hole has one event horizon, for other cases $\Delta$ has 
two zeros and consequently two horizons are present. If 
$\Delta$ has one zero, $r=r_+$, we proceed as follows. 
In the region where $\Delta<0$ we introduce the Kruskal 
coordinates $U=+e^{-k u}$ and $V=+e^{+k v}$ and so  
$UV=+e^{k(v-u)}$. In the region where $\Delta>0$ we define 
the Kruskal coordinates as $U=-e^{-k u}$ and $V=+e^{+k v}$ 
in order that $UV=-e^{k(v-u)}$. 
The signal of the product $UV$ is chosen so that the factor 
$\Delta/UV$, that appears in the metric coefficient $g_{UV}$, 
is negative. The constant $k$ is introduced in order that 
the limit of $\Delta/UV$ as $r\rightarrow r_+$ stays finite. 

If $\Delta$ has two zeros, $r=r_-$ and $r=r_+$ 
(with $r_-<r_+$), one has to introduce a Kruskal coordinate 
patch around each of the zeros of $\Delta$. The first patch 
constructed around $r_-$ is valid for $0<r<r_+$. For this 
patch, in the region where $\Delta<0$ we introduce the 
Kruskal coordinates $U=+e^{+k_{-} u}$ and $V=+e^{-k_{-} v}$ 
and so  $UV=+e^{k_{-}(u-v)}$. In the region where $\Delta>0$ 
we define the Kruskal coordinates as $U=-e^{+k_{-} u}$ and 
$V=+e^{-k_{-} v}$ in order that $UV=-e^{k_{-}(u-v)}$. The 
metric defined by this Kruskal coordinates is regular in 
the patch $0<r<r_+$ and, in particular, is regular at $r_-$. 
However, it is singular at $r_+$. To have a metric non singular 
at $r_+$ one has to define new Kruskal coordinates for the 
second patch which is constructed around $r_+$ and is valid for 
$r_-<r<\infty$. For this patch, in the region where $\Delta<0$ 
we introduce the Kruskal coordinates $U=+e^{-k_{+} u}$ and 
$V=+e^{+k_{+} v}$ and so  $UV=+e^{k_{+}(v-u)}$. In the region 
where $\Delta>0$ we define the Kruskal coordinates as 
$U=-e^{-k_{+} u}$ and $V=+e^{+k_{+} v}$ in order that 
$UV=-e^{k_{+}(v-u)}$. $k_{-}$ and $k_{+}$ are constants 
obeying the same condition defined above for $k$ and the sign 
of $UV$ is also chosen in order to have metric coefficient 
$g_{UV} \propto \Delta/UV$ negative. Now, these two different 
patches have to be joined together.
Finally, to construct the Penrose diagram one has to define 
the Penrose coordinates by the usual arctangent functions of 
$U$ and $V$: ${\cal{U}}=\arctan U$ and ${\cal{V}}=\arctan V$.

The horizon is mapped into two mutual perpendicular straight 
null lines at $45^{\rm o}$. 
In general, to find what kind of curve describes the lines 
$r=0$ or $r=\infty$ one has to take the limit of $UV$ as 
$r\rightarrow 0$, in the case of $r=0$, and the limit of $UV$ 
as $r\rightarrow \infty$, in the case of $r=\infty$. If this 
limit is $\infty$ the corresponding line is mapped into a 
curved null line. If the limit is $-1$ the corresponding line 
is mapped into a curved timelike line and finally, when the 
limit is $+1$ the line is mapped into a curved spacelike line. 
The asymptotic lines are drawn as straight lines although in 
the coordinates ${\cal{U}}$ and ${\cal{V}}$ they should be 
curved outwards, bulged. It is always possible to change 
coordinates so that the asymptotic lines are indeed 
straight lines.

The lines of infinite redshift, $r=r_{\rm rs}$, are given by 
the vanishing of the $g_{tt}$ metric component. There are 
closed timelike curves, $r_{CTC}$, whenever 
$g_{\varphi \varphi}<0$.  

The Penrose diagram for the static charged black hole is similar 
to the one drawn for the corresponding rotating  charged
 black hole. The only difference is that the infinite redshift 
lines coincide with the horizons and there are no closed 
timelike surfaces. This similarity is due to the fact that the 
rotating black hole is obtained from the static one by applying
the coordinate transformations (\ref{TRANSF_J}). 

In practice, to find the curve that describes the asymptotic 
limits $r=0$ and $r=\infty$, we can use a trick. We study the 
behavior of $\Delta$ at  the asymptotic limits, i.e. we find 
which term of $\Delta$ dominates as  $r\rightarrow 0$ and 
$r\rightarrow \infty$. Then, in the asymptotic region, we take 
$\Delta \sim \Delta_0$ in the vicinity of $r=0$ and 
$\Delta \sim \Delta_{\infty}$ in the vicinity of $r=\infty$. 
The above procedure of finding the Kruskal coordinates is 
then applied to the asymptotic regions, e.g. in the vicinity 
of $r=\infty$ we can take 
$r_{\ast} \sim r_{\ast}^{\infty}=\int(\Delta_{\infty})^{-1}dr$ 
and find the character of the $r=\infty$ curve from the limit 
of $UV$ as $r\rightarrow \infty$.

\vskip 3mm
\noindent
{\bf 6.2. Analysis of the geodesic structure}

\vskip 3mm
Let us now consider the geodesic motion.
The equations governing the geodesics can be derived from the
 Lagrangian
\begin{equation}
{\cal{L}}=\frac{1}{2}g_{\mu\nu}\frac{dx^{\mu}}{d \tau}
       \frac{dx^{\nu}}{d \tau}=-\frac{\delta}{2}\:,
                                 \label{LAG)}  \\
\end{equation}
where $\tau$ is an affine parameter along the geodesic which, 
for a timelike geodesic, can be identified with the proper 
time of the particle along the geodesic. For a null geodesic 
one has $\delta=0$ and for a timelike geodesic $\delta=+1$. 
From the Euler-Lagrange equations one gets that the generalized 
momentums associated with the time coordinate and angular 
coordinate are constants: $p_t=E$ and $p_{\varphi}=L$. The 
constant $E$ is related to the timelike Killing vector 
$(\partial/\partial t)^{\mu}$ which reflects the time 
translation invariance of the metric, while the constant 
$L$ is associated to the spacelike Killing vector 
$(\partial/\partial \varphi)^{\mu}$ which reflects the 
invariance of the metric under rotation. Note that since the 
spacetime is not asymptotically flat, the constants $E$ and 
$L$ cannot be interpreted as the local energy and angular 
momentum at infinity. 

From the geodesics equations we can derive two equations  which 
will be specially useful since they describe the behavior of 
geodesic motion along the radial coordinate. For $\omega<-2$ 
and $\omega>-1$ these are
\begin{eqnarray}
r^2 \dot{r}^2 &=& - {\biggl [}r^2 \delta 
        +\frac{(\omega+2)c_0^2}{2(\omega+1)-(2\omega +3)\Omega} 
       {\biggr ]}  \Delta +
        \frac{2(\omega+1)c_1^2}{2(\omega+1)-(2\omega +3)\Omega} 
                 r^2 
          \nonumber \\
                                 \label{GEOD_A}  \\ 
r^2 \dot{r}^2 &=& (E^2-\alpha^2L^2)r^2+\frac{M c_0^2}
    {2(\alpha r)^{\frac{1}{\omega+1}}} - 
    \frac{\sqrt{2}(\omega+1)}
   {16 \sqrt{2-\Omega}}
   \frac{c_0^2 Q^2}{(\alpha r)^{\frac{2}{\omega+1}}} 
   -r^2 \delta \Delta\:.
                                        \label{GEOD_A_2}
\end{eqnarray}
For $-2<\omega<-3/2$ and $-3/2<\omega<-1$ the two useful 
equations are
\begin{eqnarray}
r^2 \dot{r}^2 &=&   - {\biggl [}r^2 \delta 
   -\frac{(\omega+2)c_0^2}{2(\omega+1)-(2\omega +3)\Omega} 
   {\biggr ]} \Delta +
        \frac{2(\omega+1)c_1^2}{2(\omega+1)-(2\omega +3)\Omega} 
              r^2               \nonumber \\
                                 \label{GEOD_B}  \\
r^2 \dot{r}^2 &=& (E^2-\alpha^2L^2)r^2-\frac{M c_0^2}
{2 (\alpha r)^{\frac{1}{\omega+1}}} + \frac{\sqrt{2}(\omega+1)}
{16 \sqrt{2-\Omega}}
 \frac{c_0^2 Q^2}{(\alpha r)^{\frac{2}{\omega+1}}}
 -r^2 \delta \Delta \:.    
                                        \label{GEOD_B_2}
\end{eqnarray}
In the above equations $\Delta$ is the inverse of the metric 
component $g_{rr}$  and we have introduced the definitions
\begin{eqnarray}
c_0^2  &=& {\biggl [}\frac{\sqrt{|\Omega|}}{\alpha}E \mp
    \sqrt{{\biggl |}\frac{\omega +1}{\omega +2}{\biggl |}} 
    \sqrt{2-\Omega}\:L {\biggr ]}^2\:, \:\:\:  {\rm and}
                                        \label{c0}  \\ 
c_1^2 &=&  {\biggl [}\sqrt{\frac{2-\Omega}{2}}E-
    \sqrt{{\biggl |}\frac{\omega +2}{2(\omega +1)}{\biggl |}}
    \sqrt{|\Omega|} \:\alpha L {\biggr ]}^2 \:,   
                                            \label{c1}
\end{eqnarray}
where in (\ref{c0}) the minus sign is valid for $\omega<-2$ 
and $\omega>-1$ and the plus sign is applied when 
$-2<\omega<-3/2$ or $-3/2<\omega<-1$.  
There are turning points, $r_{\rm tp}$, whenever $\dot{r}=0$. 
If this  is the case, equations (\ref{GEOD_A}) and 
(\ref{GEOD_B}) will allow us to make considerations about 
the position of the turning point relatively to the position 
of the horizon. For this purpose it will be important to note 
that for $\omega<-2$ and $\omega>-1$ one has 
$0 \leq \Omega \leq 1$, and for $-2<\omega<-3/2$ or 
$-3/2<\omega<-1$ we have that $\Omega <0$. As a first and 
general example of the interest of equations (\ref{GEOD_A}) 
and (\ref{GEOD_B}) note that the turning points coincide with 
the horizons when the energy and the angular momentum are such 
that $c_1=0$. From equations (\ref{GEOD_A_2}), (\ref{GEOD_B_2}) 
and after some graphic computation we can reach interesting 
conclusions.

\vskip 3mm
\noindent
{\bf 6.3. Penrose diagrams and geodesics for each range of 
$\omega$}

\vskip 3mm

We are now in position to draw the Penrose diagrams and 
study the geodesic motion. Besides the cases $\omega=-2$ 
and $\omega=-3/2$, for each different range of $\omega$ 
($\omega<-2$, $-2<\omega<-3/2$, $-3/2<\omega<-1$, 
$\omega>-1$ and $\omega=\pm \infty$) we will consider a 
particular value of $\omega$. These will be precisely the
 ones that have been analyzed in the uncharged study of 
action (\ref{ACCAO}) \cite{Sa_Lemos_Static,Sa_Lemos_Rotat}. 

We will study the solutions with positive mass, $M>0$, and 
describe briefly the Penrose diagrams of the solutions 
with negative mass.

\vskip 3mm
{\bf 6.3.1 $\omega<-2$}
\vskip 3mm

For this range of $\omega$ there is the possibility of 
having black holes with positive mass whenever 
$|\alpha J|\leq\frac{|2\omega+3|M}{2\sqrt{(\omega+1)(\omega+2)}}$.
We are going to analyze the typical case $\omega=-3$. 
The $\Delta$ function, (\ref{DELTA_TODOS}), is 
\begin{equation}
 \Delta =(\alpha r)^2 -b\sqrt{\alpha r}+c(\alpha r)\:,
                                 \label{DEL_(-3)}  \\
\end{equation}
where from (\ref{b}) and (\ref{c^2}) one has that $b>0$ and 
$c\equiv k\chi^2<0$ (if $M>0$). For 
$|\alpha J|\leq\frac{3M}{2\sqrt{2}}$, $\Delta$ has always one and 
only one zero given by
\begin{eqnarray}
& & r_+=\frac{1}{3\alpha}{\biggl [}-2c +\frac{c^2}{s}
    +s{\biggr ]}\:, \:\:\:\:\:{\rm where}  \nonumber \\
& & s={\biggl [}\frac{1}{2}{\biggl (} 27b^2+2c^3+3\sqrt{3}b 
    \sqrt{27b^2+4c^3}{\biggr )}{\biggr ]}^{\frac{1}{3}}\:,
                                 \label{Zero_(-3)}
\end{eqnarray}
$\Delta>0$ for $r>r_+$ and $\Delta<0$ for $r<r_+$. 
The curvature singularity at $r=0$ is a spacelike 
line in the Penrose diagram while $r=+\infty$ is a timelike line.  
The Penrose diagram is drawn in figure 1. There is no extreme
black hole for this case.
\vskip 3mm 
\centerline{\epsffile
{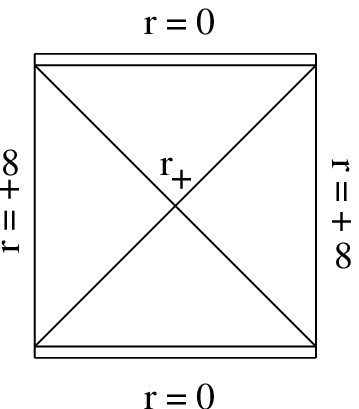}}
\vskip 1mm
{\small
{\bf Figure 1:} Penrose diagram for the $\omega=-3$, 
$M>0$, $|\alpha J|\leq 3M/2\sqrt{2}$ black hole.
}
\vskip 3mm

When we consider the solutions with negative mass  we conclude 
that, for $|\alpha J|\leq\frac{3|M|}{2\sqrt{2}}$, one has black holes 
with two horizons, with one (extreme case) or a spacetime without 
black holes. For the black hole with two horizons the Penrose 
diagram is shown in figure 2.(a) of and the extreme black hole has a 
Penrose diagram which is  drawn in figure 2.(b).
\vskip 3mm
\centerline{\epsffile
  {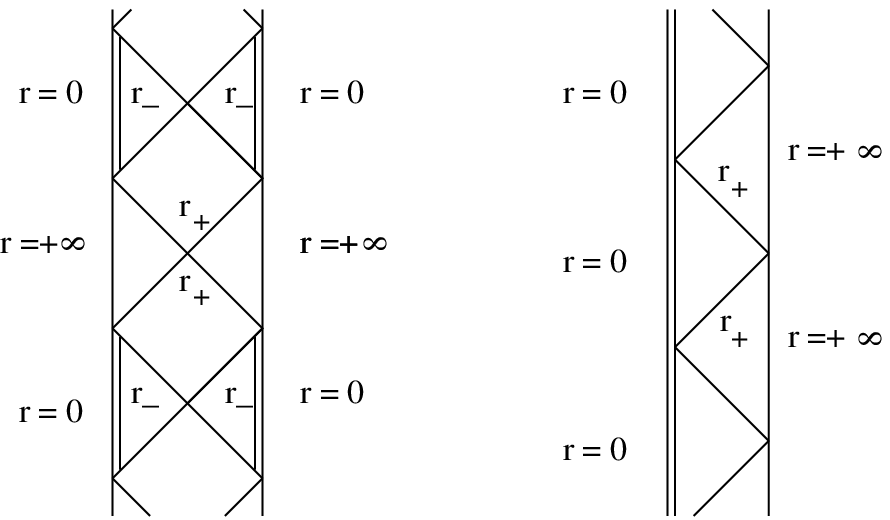}}
\vskip 1mm
{\small
\centerline{\noindent 
{\bf (a)} \hskip 5cm {\bf (b)}}

{\bf Figure 2: (a)} Penrose diagram for the $\omega=-3$, $M<0$, 
$|\alpha J|\leq 3|M|/2\sqrt{2}$ black hole with two horizons. 
{\bf (b)} Penrose diagram for the $\omega=-3$, $M<0$, 
$|\alpha J|\leq 3|M|/2\sqrt{2}$ extreme black hole.
}
\vskip 3mm

Let us now consider the geodesic motion. Analyzing (\ref{GEOD_A}) 
and noting that $0 \leq \Omega \leq 1$ we see that for the null and 
timelike geodesics the coefficient of $\Delta$ is always negative so, 
for $0<r<r_+$, the first term of (\ref{GEOD_A}) is positive. Since 
the second term is positive or null we conclude that whenever there 
are turning points, they are $r_{\rm tp}^1=0$ and 
$r_{\rm tp}^2\geq r_+$. From (\ref{GEOD_A_2}) we conclude the 
following about the geodesic motion. (i) If $E^2-\alpha^2L^2 \geq 0$ 
null particles produced at $r=0$ escape to $r=+\infty$ and null 
particles coming in from infinity are scattered at $r_{\rm tp}^1=0$ 
and spiral back to infinity. (ii) Null geodesics with $E^2-\alpha^2L^2< 0$ 
are bounded between the singularity $r_{\rm tp}^1=0$ and a maximum 
($r_{\rm tp}^2\geq r_+$) radial distance. The turning point 
$r_{\rm tp}^2$ is exactly at the horizon $r_+$ if and only if the 
energy and the angular momentum are such that $c_1=0$. (iii) Null 
geodesics with energy and angular momentum such that $c_0=0$ can 
reach and ``stay'' at the curvature singularity $r=0$. (iv) All the 
timelike geodesics present the same features as the null geodesics with 
$E^2-\alpha^2L^2< 0$. So, any timelike geodesic is bounded  within 
the region $r_{\rm tp}^1\leq r \leq r_{\rm tp}^2$ (with $r_{\rm tp}^1=0$ 
and $r_{\rm tp}^2\geq r_+$), and no timelike particle can either escape 
to infinity or reach and ``stay'' at $r=0$. (v) Neither null or timelike 
geodesics have stable or unstable circular orbits.

\vskip 15mm
{\bf 6.3.2 $\omega=-2$}
\vskip 3mm

For $\omega=-2$, $\Delta$ is given by (\ref{DELTA-2}) and, in the case $M>0$,
$\Delta$ has one zero given by 
\begin{equation}
r_+=b {\biggl [} 4\chi^2{\rm ProdLog}{\biggl (} 
\frac{be^{\frac{1}{4\chi^2}}}{4\chi^2}{\biggr )}{\biggr ]}^{-1},
                                 \label{ZERO_(-2)}  \\
\end{equation}
with ${\rm ProdLog}(x)=z$ being such that $x=ze^z$.
The scalar $R_{\mu\nu}R^{\mu\nu}$ (\ref{R-2})  
diverge at $r=0$ and $r=+\infty$. 
For $|J|>|M|$, $\Delta$ is positive for $r>r_+$  and negative for $r<r_+$.  
At $r=+\infty$ the curvature singularity is timelike 
while $r=0$ is a null curvature singularity.  The Penrose 
diagram is represented in figure 3.(a). 
For $|J|<|M|$, unlike the uncharged case, the theory also 
has a black hole with a horizon located at $r_+$ given by 
(\ref{ZERO_(-2)}). $\Delta$ is negative for $r>r_+$  and 
positive for $r<r_+$.  
The $r=+\infty$ curvature singularity is spacelike and 
$r=0$ is a null curvature singularity.  The Penrose diagram 
is drawn in figure 3.(b). 
For $|J|=|M|$ one has $b= \chi=0$. Then, $r=0$ is a naked null 
singularity and the boundary $r=\infty$ changes character and 
has no singularity, being a timelike line in the Penrose diagram 
drawn in figure 3.(c).
\vskip 3mm
\centerline{\epsffile
{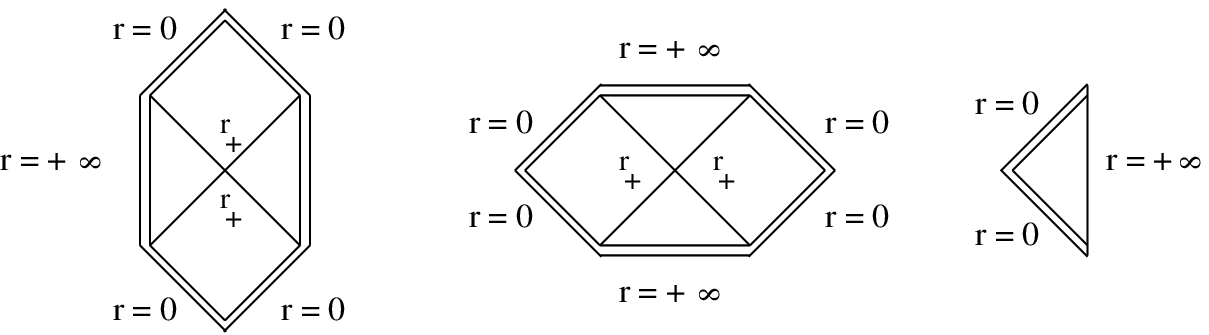}}
\vskip 1mm
{\small
\centerline{\noindent \hskip 0.5cm {\bf (a)} \hskip 4cm {\bf (b)} 
\hskip 3cm {\bf (c)}}

{\bf Figure 3: (a)} Penrose diagram for the black 
hole of:
i) $\omega=-2$, $M>0$, $|J|>M$; 
ii) $\omega=-9/5$, $M<0$.
{\bf (b)} Penrose diagram for the  $\omega=-2$, $M>0$, 
$|J|<M$ naked singularity. 
{\bf (c)} Penrose diagram for the  $\omega=-2$, $M>0$, 
$|J|=M$ naked singularity.
}
\vskip 3mm

Now, we study the geodesic motion for $M>0$.  
The behavior of geodesic motion along the radial coordinate can 
be obtained from the following two equations
\begin{eqnarray}
r^2 \dot{r}^2 &=& - {\bigl [} r^2\delta +c_0^2 {\bigr ]} 
\Delta +c_1^2r^2   
                                \label{GEOD_(-2)}  \\ 
r^2\dot{r}^2 &=& (E^2-L^2)r^2-\frac{\chi^2c_0^2}{4}r^2 \ln r 
 +\frac{J^2-M^2}{M} c_0^2 r 
   -r^2 \delta \Delta\:,
                                        \label{GEOD_(-2)_2}
\end{eqnarray}
where
\begin{eqnarray}
c_0^2 &=& {\biggl [}{\biggl (}\frac{J^2}{M^2}-1 {\biggl )}^{-1/2}E
          - {\biggl (}1-\frac{M^2}{J^2} {\biggl )}^{-1/2} L   
{\biggr ]}^2 \:,                           \label{GEOD_(-2)_c0}  \\ 
c_1^2 &=& {\biggl [}{\biggl (}1-\frac{M^2}{J^2} {\biggl )}^{-1/2}E
          - {\biggl (}\frac{J^2}{M^2}-1 {\biggl )}^{-1/2} L   
{\biggr ]}^2 \:.
                                        \label{GEOD_(-2)_2_c1}
\end{eqnarray} 
We first consider the case $|J|>M$. From equation 
(\ref{GEOD_(-2)}) we conclude that whenever there are 
turning points, they are given by $r_{\rm tp}^1=0$ and 
$r_{\rm tp}^2\geq r_+$. The turning point $r_{\rm tp}^2$ is 
exactly at the horizon $r_+$ if and only if the energy and 
the angular momentum are such that $c_1=0$, i.e,  $E=M L/J$.
 From the graphic computation of (\ref{GEOD_(-2)_2}) we 
conclude that: (i)
the only particles that can escape to $r=+\infty$ or $r=0$ 
are null particles that satisfy $c_0=0$ which implies 
$E=J L/M$; (ii) all other null geodesics and all timelike 
geodesics are bounded between $r_{\rm tp}^1=0$ and a maximum 
($r_{\rm tp}^2\geq r_+$) radial distance. 

 For $|J|<M$ one has that: (i) timelike and null spiraling 
particles with  $E \neq J L/M$ start at $r_{\rm tp}\geq r_+$ 
and reach infinity radial distances or timelike and null 
geodesics start at $r=+\infty$ and spiral toward 
$r_{\rm tp}$ and then return back to infinity;
(ii)  null particles can escape to $r=+\infty$ or $r=0$ if 
$c_0=0$, i.e., $E=J L/M$; (iii) timelike geodesics with 
$E=J L/M$ can be bounded  between $r_{\rm tp}^1=0$ and a 
maximum ($r_{\rm tp}^2\geq r_+$) radial distance, or  
 start at $r_{\rm tp}^3 \geq r_+$ and reach infinity radial 
distances.

\vskip 3mm
{\bf 6.3.3 $-2<\omega<-3/2$}
\vskip 3mm

In this range of $\omega$, spacetime has no black holes since 
from (\ref{b}) and (\ref{c^2}) one has $b<0$ and $k \chi^2>0$ 
so the $\Delta$ function (\ref{DELTA_TODOS}) is always positive. 
The curvature singularity $r=0$ is a naked null singularity and 
the curvature singularity $r=+\infty$ is a naked timelike 
singularity.  The Penrose diagram is drawn in figure 4.

When we consider the solutions with negative mass  we conclude 
that one might have black holes with one horizon. If this is 
the case, the Penrose diagram is exactly equal to the one 
shown in figure 3.(a) (which represents the typical case 
$\omega=-9/5$). 
\vskip 3mm
\centerline{\epsffile
{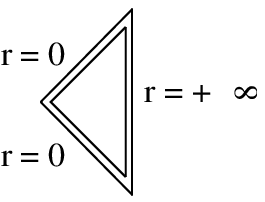}}
\vskip 1mm
{\small
{\bf Figure 4:} Penrose diagram for the spacetime of: 
i) $\omega=-2$, $M>0$, $|J|=M$; 
ii) $-2<\omega<-3/2$, $M>0$; 
iii) $\omega=-3/2$ (large $\Lambda/\chi^2$);
iv) $\omega=-4/3$, $M<0$, $|\alpha J|=M$ 
(the only difference is that $r=0$ is a topological singularity rather 
than a curvature singularity).
}
\vskip 3mm

Although there are no black holes with positive mass, it 
is interesting to study the geodesics that we might expect 
for this spacetime. From graphic computation of  
(\ref{GEOD_B_2}) we conclude the following for null and 
timelike geodesics. (i) For $E^2-\alpha^2L^2 \leq 0$ there 
is no possible motion. (ii) This situation also occurs for 
small positive values of $E^2-\alpha^2L^2$. (iii) However, 
when we increase the positive value of $E^2-\alpha^2L^2$ 
there is a critical value for which a stable circular 
orbit is allowed. (iii) And for positive values of 
$E^2-\alpha^2L^2$ above the critical value, null and 
timelike geodesics are bounded between a minimum 
($r_{\rm tp}^1$) and a maximum ($r_{\rm tp}^2$) radial 
distance. (iv)  
Null particles with energy and angular momentum satisfying 
$c_0=0$ that are produced at $r=0$ escape to $r=+\infty$ 
and null particles coming in from infinity are scattered 
at $r_{\rm tp}=0$ and spiral back to infinity.
(vi) Timelike particles with energy and angular momentum 
satisfying $c_0=0$ are bounded  within the region 
$0\leq r \leq r_{\rm tp}$, with $r_{\rm tp}$ finite.

\vskip 3mm
{\bf 6.3.4 $\omega=-3/2$}
\vskip 3mm

In this case, $\Delta$ is given by (\ref{DELTA-3/2}).
Depending on the value of $\frac{\Lambda}{\chi^2}$ one has 
black holes with two horizons (small $\frac{\Lambda}{\chi^2}$), 
with one (extreme case) or a spacetime without black holes 
(large $\frac{\Lambda}{\chi^2}$).

For the black hole with two horizons, we have $\Delta>0$ in 
$r<r_-$ and $r>r_+$. For the patch
$r_-<r<\infty$, the curvature singularity $r=\infty$ is mapped 
into two symmetrical timelike lines and the horizon $r=r_+$ 
is mapped into two mutual perpendicular straight lines at 
$45^{\rm o}$. For the patch
$0<r<r_+$, the curvature singularity $r=0$ is mapped into a pair 
of two null lines and the horizon $r=r_-$ is mapped into two 
mutual perpendicular straight lines at $45^{\rm o}$. One has 
to join these two different patches and then repeat them over 
in the vertical. The resulting Penrose diagram is shown in 
figure 5.(a).
For the extreme black hole the two curvature singularities 
$r=0$  and 
$r=\infty$ are still null and timelike lines (respectively), 
but the event and inner horizon join together in a single 
horizon $r_+$. The Penrose diagram is like the one drawn in 
figure 5.(b). 
The spacetime with no black hole present has a Penrose diagram 
like the one represented in figure 4.
\vskip 3mm
\centerline{\epsffile
  {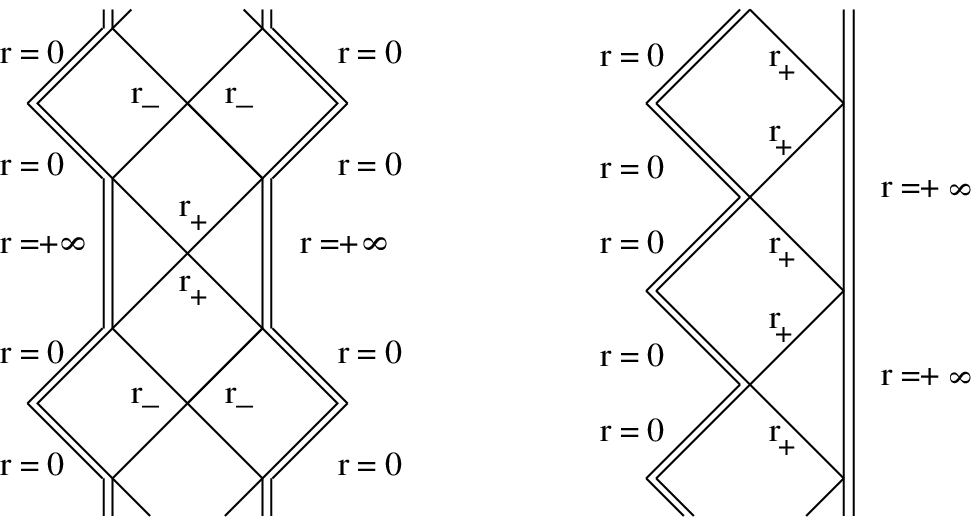}}
{\small
\centerline{\noindent {\bf (a)} \hskip 5cm {\bf (b)}}
\vskip 1mm

{\bf Figure 5: (a)} Penrose diagram for the black hole with two 
horizons of: 
i) the $\omega=-3/2$, (small $\Lambda/\chi^2$); 
ii)  $\omega=-4/3$, $M<0$, $|\alpha J|< |M|$ (the only 
difference is that $r=0$ is a topological singularity rather 
than a curvature 
singularity).

{\bf (b)} Penrose diagram for the extreme black 
hole of:
i) $\omega=-3/2$;
ii) $\omega=-4/3$, $M<0$, $|\alpha J|< |M|$
(the only difference is that $r=0$ is a topological singularity rather 
than a curvature 
singularity).
}
\vskip 1mm

Now, we study the geodesic motion.  
   
The behavior of geodesic motion along the radial coordinate 
can be obtained from the following two equations
\begin{eqnarray}
r^2 \dot{r}^2 &=& - {\bigl [}r^2 \delta +c_0^2 {\bigr ]} 
\Delta +c_0^2 r^2  
                                                 \label{GEOD_(-3/2)}  \\ 
\dot{r}^2 &=& c_0^2 {\bigl [}\Lambda \ln(br)-\chi^2r^2+1{\bigr ]} 
   - \delta \Delta\:,
                                        \label{GEOD_(-3/2)_2}
\end{eqnarray} 
where $c_0=(\theta E-\gamma L)$. From equation (\ref{GEOD_(-3/2)}) we 
conclude that whenever there are turning points, they are given by 
$r_{\rm tp}^1 \leq r_-$ and $r_{\rm tp}^2 \geq r_+$. The turning 
points coincide with the horizons if and only if the energy and the 
angular momentum are such that $c_0=0$.
From the graphic computation of (\ref{GEOD_(-3/2)_2}) we conclude that:
(i) Whenever $c_0=0$ null particles describe stable circular 
orbits wherever they are located; (ii) Null geodesics with 
$c_0 \neq 0$ and all timelike geodesics describe a bound orbit 
between $r_{\rm tp}^1$ and $r_{\rm tp}^2$. 

For the extreme black hole the two horizons coincide so all null 
geodesics  and all timelike geodesics describe a stable circular orbit.

\vskip 10mm
{\bf 6.3.5 $-3/2<\omega<-1$}
\vskip 3mm 

 For this range of $\omega$ there is the possibility of having 
black holes with positive mass whenever $|\alpha J|>M$ (Table 1).
We are going to analyze the typical case $\omega=-4/3$. 
The $\Delta$ function, (\ref{DELTA_TODOS}), is 
\begin{equation} 
\Delta =(\alpha r)^2 -b(\alpha r)^3+c(\alpha r)^6\:,
                                 \label{DEL_(-3/2)_(-1)}  \\
\end{equation}
where from (\ref{b}) and (\ref{c^2}) one has that $b>0$ and 
$c\equiv k\chi^2<0$ if $M>0$ and $|\alpha J|>M$.
$\Delta$ is negative for $r>r_+$ and positive for $r<r_+$, 
where $r_+$ is the only zero of $\Delta$ given by
\begin{eqnarray}
& & r_+=\frac{1}{2\alpha} {\biggl (}\frac{b}{c} {\biggl )}^{1/3} 
{\biggl [}\sqrt{s} -\sqrt{\frac{2}{\sqrt{s}}-s}{\biggr ]}\:, 
\:\:\:\:\:{\rm where}  \nonumber \\
& & s={\biggl [}\frac{1}{2}+\frac{1}{2}\sqrt{1-\frac{4^4 c}
{3^3 b^4} }{\biggr ]}^{\frac{1}{3}}+{\biggl [}\frac{1}{2}-
\frac{1}{2}\sqrt{1-\frac{4^4 c}{3^3 b^4} }{\biggr ]}^{\frac{1}{3}}\:.  
\label{Zero_(-4/3)}
\end{eqnarray}

The physical curvature singularity is located inside the horizon at
$r=+\infty$ and is a spacelike line in the Penrose diagram. 
At $r=0$ there is a null topological singularity.  The Penrose 
diagram is sketched in figure 6.

\vskip 3mm
\centerline{\epsffile
  {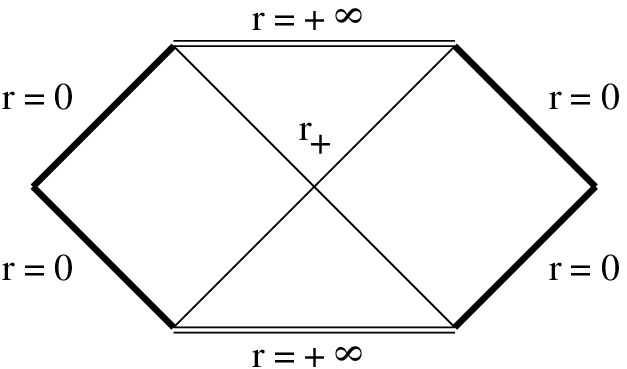}}
\vskip 1mm
{\small
{\bf Figure 6:} Penrose diagram for the black hole of: 
i)  $\omega=-4/3$, $M>0$, $|\alpha J|> M$;
ii)  $\omega=-4/3$, $M<0$, $|\alpha J|> |M|$.
}
\vskip 3mm 

When we consider the solutions with negative mass  we conclude that, for 
$|\alpha J|>|M|$, one has black holes with a Penrose diagram 
equal to the one drawn in figure 6. For $|\alpha J|<|M|$ one 
has black holes with two horizons, with one (extreme case) 
or a spacetime without black holes. For the black hole with 
two horizons the Penrose diagram is similar to the one shown 
in figure 5.(a)  and the extreme black hole has a Penrose 
diagram which is similar to the one drawn in figure 5.(b). 
For $|\alpha J|=|M|$ the Penrose diagram is similar to figure 4. 
The only difference is the fact that $r=0$ is now a topological 
singularity rather than a curvature singularity.

Let us now consider the geodesic motion for positive mass. 
Analyzing (\ref{GEOD_B}) and noting that from (\ref{OMEGA}) 
the condition $|\alpha J|>M$ implies 
$\Omega<-2$ we conclude that, for null geodesics, whenever there 
are turning points, they are $r_{\rm tp}^1=0$ and 
$r_{\rm tp}^2\geq r_+$. From (\ref{GEOD_B_2}) we conclude 
the following about the null geodesic motion. 
(i) For $E^2-\alpha^2L^2 \leq 0$ there is no possible motion.
(ii)  Null geodesics with $E^2-\alpha^2L^2> 0$ are bounded 
between the singularity $r_{\rm tp}^1=0$ and a maximum 
($r_{\rm tp}^2\geq r_+$) radial distance. The turning 
point $r_{\rm tp}^2$ is exactly at the horizon $r_+$ if 
and only if the energy and the angular momentum are such 
that $c_1=0$.

The timelike geodesic motion is radically different. (i) 
For $E^2-\alpha^2L^2 \leq 0$ we have timelike spiraling 
particles that start at $r_{\rm tp}$ and reach infinity 
radial distances or timelike geodesics that start at 
$r=+\infty$ and spiral toward $r_{\rm tp}$ and then return 
back to infinity. (ii) For small positive values of 
$E^2-\alpha^2L^2$, timelike particles that are produced 
at $r=0$ escape to $r=+\infty$ and timelike particles 
coming in from infinity are scattered at $r_{\rm tp}=0$ 
and spiral back to infinity.
(iii) When we increase the positive value of $E^2-\alpha^2L^2$ 
there is a critical value for which an unstable circular 
orbit is allowed. (iv) And for positive values of 
$E^2-\alpha^2L^2$ above the critical value,  timelike 
particles are allowed to be bounded between 
$r_{\rm tp}^1=0$ and a maximum ($r_{\rm tp}^2$) radial 
distance or to start at $r_{\rm tp}^3>r_{\rm tp}^2$ and 
escape to infinity.

\vskip 3mm
{\bf 6.3.6 $\omega>-1$}
\vskip 3mm

The range $\omega>-1$ is not discussed here since the 
properties of the typical case $\omega=0$ have been 
presented in \cite{Zanchin_Lemos}, where the three-dimensional 
gravity theory of $\omega=0$ was obtained through dimensional 
reduction from four-dimensional General Relativity with one 
Killing vector field. 

\vskip 3mm
{\bf 6.3.7 $\omega=\pm \infty$}
\vskip 3mm

The case $\omega=\pm \infty$ is also not  discussed here since 
this case reduces to the electrically charged BTZ black hole 
which has been studied in detail in \cite{BTZ_Q,PY}.  

\vskip 1cm

\noindent
{\bf 7. Hawking temperature of the charged black holes}

\vskip 3mm

To compute the Hawking temperature of the rotating black holes, 
one starts 
by writing the metric in the canonical form (\ref{MET_CANON}). 
To proceed it is necessary to first perform the coordinate 
transformation to coordinates $t$, $\tilde{\varphi}$ which corotate 
with the black hole. In other words, the angular coordinate $\varphi$ must 
be changed to  $\tilde{\varphi}$ defined by 
$\tilde{\varphi}=\varphi-\Omega_H t$, where 
$\Omega_H=-\frac{g_{t \varphi}}{g_{\varphi \varphi}}
{\bigl |}_{r=r_+}=-N^{\varphi}(r_+)$ is the angular velocity of
the black hole. With this transformation the metric 
(\ref{MET_CANON}) becomes
\begin{equation}
     ds^2 = - (N^0)^2 dt^2
            + \frac{dr^2}{f^2}
            + H^2{\biggl [}d\tilde{\varphi}+{\biggl (}N^{\varphi}(r)
             -N^{\varphi}(r_+){\biggr )}dt{\biggr ]}^2 \:,
                               \label{MET_CANON_TEMP}
\end{equation}
Then, one applies the Wick rotation $t \rightarrow -i \tau$ in 
order to obtain the euclidean counterpart of (\ref{MET_CANON_TEMP}).
Now, one studies the behavior of the euclidean metric in the 
vicinity of the event horizon, $r_+$. In this vicinity, one can 
write $N^{\varphi}(r)-N^{\varphi}(r_+) \sim 0$ and take the expansion 
$\Delta(r) \sim \frac{d\Delta(r_+)}{d r}(r-r_+)+\cdots\,$. 
One proceeds applying the variable change 
$\frac{1}{\Delta(r)}dr^2=d \rho^2$ so that  one has  
$\rho=2[\sqrt{\frac{r-r_+}{d \Delta(r_+)/d r}}$ 
 and $\Delta(\rho) \sim [d \Delta(r_+)/dr]^2 \rho^2 /4$. With this 
procedure the euclidean metric in the vicinity of the event horizon 
can be cast in the form 
$ds^2 \sim (2 \pi / \beta_H)^2 \rho^2 d \tau^2
+d \rho^2+ H^2 d\tilde{\varphi}^2$. Applying a final variable change, 
$\bar{\tau}=(2 \pi / \beta_H) \tau$, the metric becomes
$ds^2 \sim \rho^2 d \bar{\tau}^2+d \rho^2+H^2 d\tilde{\varphi}^2$. 
To avoid the canonical singularity at the event horizon one must 
demand that the period of $\bar{\tau}$ is $2 \pi$ which implies 
$0 \leq \tau \leq \beta_H$. Finally, the Hawking temperature is 
defined as $T_H=(\beta_H)^{-1}$.

Applying the above procedure, one finds for the Hawking temperature 
of the rotating black holes the following expressions
\begin{eqnarray} 
T_H &=& \frac{1}{4 \pi}\sqrt{\frac{r_+^2 {\biggl [} 2 
\alpha r_+ +\frac{b}{\omega+1}(\alpha r_+)^
{-\frac{\omega+2}{\omega+1}} 
-\frac{2k \chi^2}{\omega+1}(\alpha r_+)^{-\frac{\omega+3}
{\omega+1}}{\biggr ]}^2}{r_+^2-\frac{\theta^2}{\alpha^4} 
{\biggl [}-b (\alpha r_+)^{-\frac{1}{\omega+1}}+ k 
\chi^2(\alpha r_+)^{-\frac{2}{\omega+1}} {\biggr ]}}}  \:, 
  \nonumber \\
         & &     \hskip 5cm \omega \neq -2,-\frac32,-1 \:,
                                 \label{TEMP_TODOS}  \\
T_H &=& \frac{1}{4 \pi}\sqrt{\frac{r_+^2 {\biggl [} 
\frac{r_+}{2}{\biggl (} 4+\frac{\chi^2}{2}+ \chi^2\ln r_
+{\biggl )}-b{\biggr ]}^2}{{\biggl [}1-\frac{1}{4}\theta^2 
\chi^2\ln r_+{\biggr ]}r_+^2+\theta^2 b r_+}}  \:, 
                           \:\:\:\:\:\:\:   \omega=-2   \:,
                                 \label{TEMP_(-2)} \\
T_H &=& \frac{1}{4 \pi}\sqrt{\frac{r_+^2 
{\biggl [}  2\Lambda r_+ \ln(b r_+)+ \Lambda b r_+ 
-\frac{\chi^2r_+^3}{4}{\biggr ]}^2}
{ \Lambda \theta^2 r_+^2 \ln(b r_+)+ \gamma^2 r_+^2 
-\frac{\theta^2 \chi^2r_+^4}{16}}}  \:,   \nonumber \\
         & &     \hskip 6cm \omega=-3/2 \:,
                                 \label{TEMP_(-3/2)} 
\end{eqnarray}
where in (\ref{TEMP_TODOS}), $b$, $k \chi^2$ and $\theta^2$ 
are given by (\ref{b}), (\ref{c^2}) and (\ref{DUAS}), respectively. 

The Hawking temperature of the static charged black holes can 
be obtained from (\ref{TEMP_TODOS})-(\ref{TEMP_(-3/2)}) by taking 
$\gamma=1$ and $\theta=0$ [see (\ref{TRANSF_J})]
(see also \cite{Lemos} for the $\omega=0$ uncharged black hole).
\vskip 1cm
\noindent
{\bf 8. Conclusions}

\vskip 3mm
We have added the Maxwell action to the action of a generalized
3D dilaton gravity specified by the Brans-Dicke parameter $\omega$ 
introduced in \cite{Sa_Lemos_Static,Sa_Lemos_Rotat}.
We have concluded that for the static spacetime the electric and 
magnetic fields cannot be simultaneously  non-zero, i.e. there is 
no static dyonic solution.
In this work we have considered the electrically charged case alone.
We have found the static and rotating black hole solutions of this theory. 
It contains eight different cases that appear naturally from the 
solutions. For $\omega=0$ one gets a theory related 
(through dimensional reduction) to electrically charged four 
dimensional general relativity with one Killing vector 
\cite{Zanchin_Lemos} and for $\omega=\pm \infty$ one obtains 
electrically charged three dimensional general relativity 
\cite{CL1,BTZ_Q}. For $\omega=-1$ the Maxwell term does not alter 
the uncharged solution, i.e. one still gets the two-dimensional 
string theory black hole cross $S^1$ (For a discussion see 
\cite{Sa_Lemos_Static,Witt,MSW}). 
The uncharged theory imposed, in the case $\omega=-2$, that the 
cosmological constant had to be null. This no longer happens in 
the charged extension of the theory which imposes, for  
$\omega=-2$, a relation between the charge and the cosmological constant.

For $\omega>-3/2$ the ADM mass and angular momentum of the 
solutions are finite, well-behaved and equal to the ADM masses 
of the uncharged solutions. However, for $\omega<-3/2$ the 
ADM mass and angular momentum of the solutions have terms 
proportional to the charge that diverge at the asymptotic limit, 
as frequently occurs in the extended theories including a 
Maxwell field (see, e.g. \cite{CL1,BTZ_Q,KK2}). We have shown 
how to treat this problem. For each range of $\omega$ we have 
determined what conditions must be imposed on the ADM masses of 
the solutions in order to be possible the existence of  black holes.
Our results show that there is no upper bound on the electric charge.

The causal and geodesic structure of the charged solutions is, 
in some of the cases, quite different from the ones of the 
uncharged case. The Hawking temperature has been computed.
 
\vskip .5cm

\section*{Acknowledgments} This work was partially funded
by Funda\c c\~ao para a Ci\^encia e Tecnologia (FCT) through 
project PESO/PRO/2000/4014. O. J. C. D. 
also acknowledges finantial support from the portuguese FCT 
through PRAXIS XXI programme. J. P. S. L. thanks Observat\'orio 
Nacional do Rio de Janeiro for hospitality.

\vskip 1cm

\end{document}